\documentclass[conference]{IEEEtran}
\IEEEoverridecommandlockouts
\usepackage{cite}
\usepackage{amsmath,amssymb,amsfonts}
\usepackage{algorithmic}
\usepackage{graphicx}
\usepackage{textcomp}
\usepackage{xcolor}
\usepackage{soul}
\usepackage{subfig}   % in preamble
\usepackage{graphicx} % in preamble
\usepackage{float}    % optional
\usepackage{placeins}
\def\BibTeX{{\rm B\kern-.05em{\sc i\kern-.025em b}\kern-.08em
    T\kern-.1667em\lower.7ex\hbox{E}\kern-.125emX}}
\begin{document}

\title{Beyond the Interface: Redefining UX for Society-in-the-Loop AI Systems}
\author{\IEEEauthorblockN{1\textsuperscript{st} Nahal Mafi}
\IEEEauthorblockA{\textit{Computer Science} \\
\textit{University of North Carolina at Charlotte (UNC Charlotte)}\\
Charlotte, USA \\
nmafi@charlotte.edu}
\and
\IEEEauthorblockN{2\textsuperscript{nd} Sahar Maleki}
\IEEEauthorblockA{\textit{Belk College of Business} \\
\textit{UNC Charlotte}\\
Charlotte, USA \\
smaleki1@charlotte.edu}
\and
\IEEEauthorblockN{3\textsuperscript{rd} Babak Rahimi~Ardabili}
\IEEEauthorblockA{\textit{Electrical and Computer Engineering} \\
\textit{UNC Charlotte}\\
Charlotte, USA \\
brahimia@charlotte.edu}
\and
\IEEEauthorblockN{4\textsuperscript{th} Hamed Tabkhi}
\IEEEauthorblockA{\textit{Electrical and Computer Engineering} \\
\textit{UNC Charlotte}\\
Charlotte, USA \\
htabkhiv@charlotte.edu}
}

\maketitle

\begin{abstract}

Artificial intelligence systems increasingly operate in decision-critical environments where probabilistic outputs and Human-in-the-Loop (HITL) interactions reshape user engagement. Traditional user experience (UX) frameworks, designed for deterministic systems, fail to capture these evolving sociotechnical dynamics. This paper argues that in AI-enabled HITL systems, UX must transcend frontend usability to encompass backend performance, organizational workflows, and decision-making structures.

We employ a mixed-methods approach, combining an inductive social construction analysis of 269 stakeholder insights with the deployment of an operational HITL video anomaly detection system. Our findings reveal that stakeholders experience AI through multifaceted themes: risk, governance, and organizational capacity. Experimental results further demonstrate how detection behavior and alert routing directly calibrate human oversight and workload. Grounded in these results, we formalize a new evaluative framework centered on four sociotechnical metrics: Accuracy (FPR/FNR), Operational Latency (response time), Adaptation Time (deployment burden), and Trust (validated automation scales). This framework redefines UX as a multi-layered construct spanning infrastructure and governance, providing a rigorous foundation for evaluating AI systems embedded within complex real-world ecosystems.

\end{abstract}

\begin{IEEEkeywords}
Human-in-the-Loop, User Experience, Sociotechnical Systems, AI Evaluation, Trust in Automation, Post-AI UX
\end{IEEEkeywords}

\section{Introduction} 
\begin{figure*}[!t]
    \centering
    \subfloat[Pre-AI UX Design Flowchart.]{%
        \includegraphics[width=0.4\linewidth, trim=1cm 3cm 4cm 3cm]{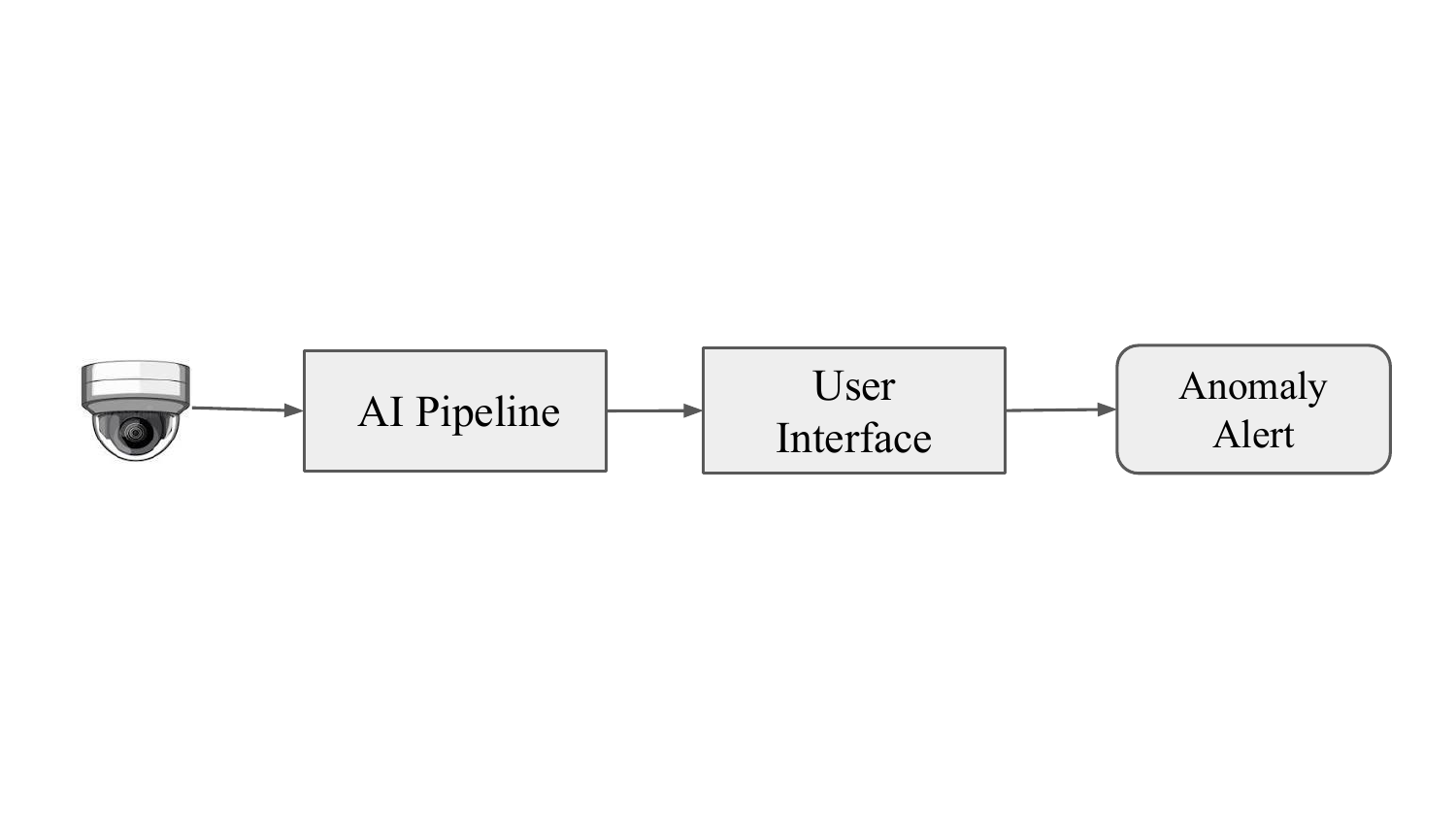}
    }
    \hfill
    \subfloat[Post-AI UX Design Flowchart.]{%
        \includegraphics[width=0.49\linewidth, trim=0.1cm 2cm 0.1cm 1cm]{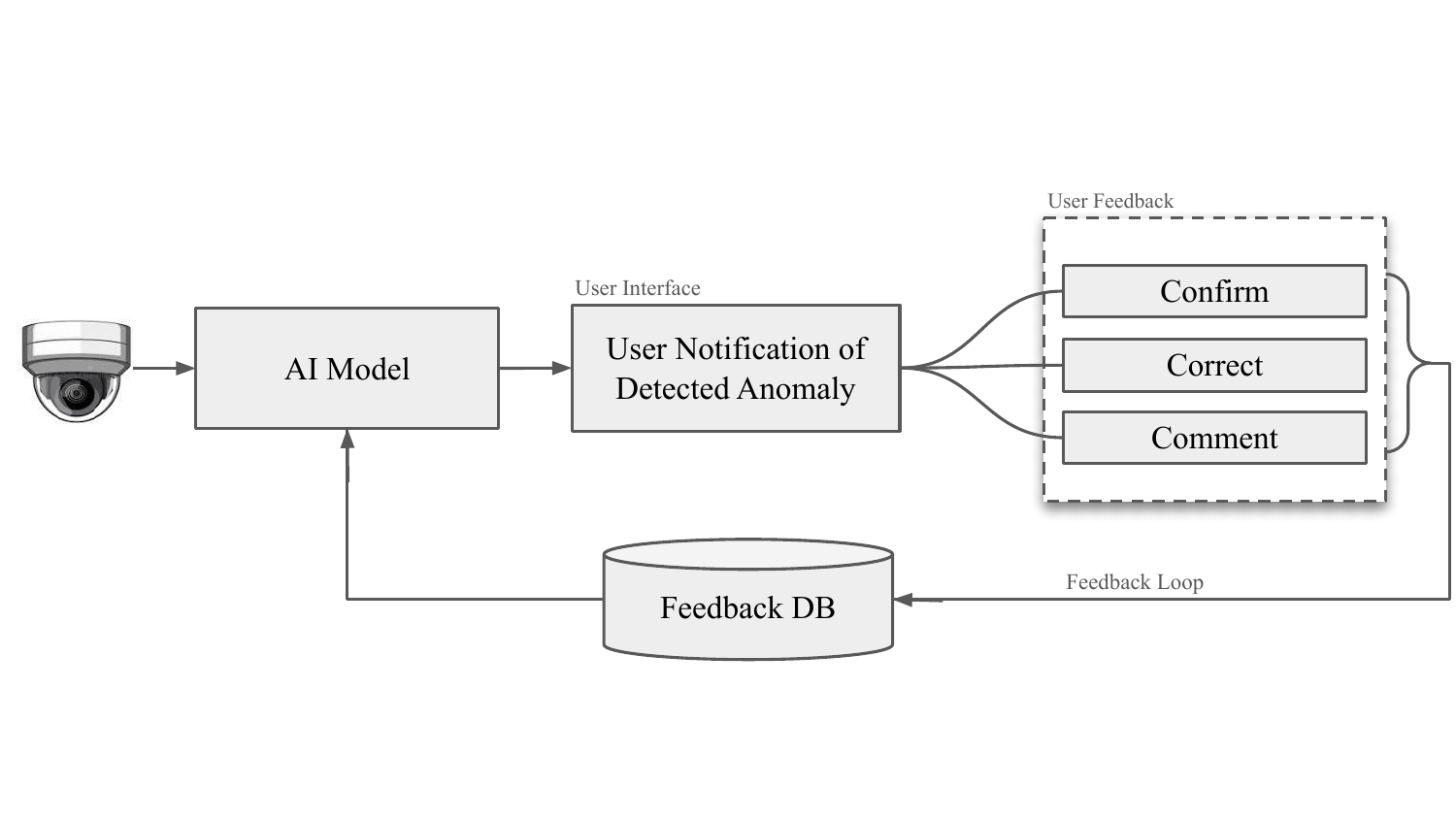}
    }
    \caption{Comparison of pre-AI (left) and post-AI (right) UX design workflows. 
The pre-AI flow represents a linear, user-driven interaction model, while the post-AI flow introduces AI-mediated detection and alerting with human-in-the-loop feedback cycles (confirm, correct, comment), transforming UX into a dynamic co-adaptive process between users and intelligent systems.}
    \label{fig:pre_post_loop}

\end{figure*}
Artificial intelligence systems are increasingly deployed in high-risk environments where their outputs directly influence operational and business outcomes. Despite their rapid adoption, these systems inherently generate probabilistic predictions that may behave inconsistently across different contexts. Because these systems are not fully deterministic, their outputs are often neither transparent nor easily interpretable \cite{b1}. For decades, traditional user experience (UX) design focused primarily on task completion, efficiency, and the reduction of interaction friction. However, these objectives are no longer sufficient in AI-driven environments.

In the emerging post-AI paradigm, users are no longer mere operators executing predefined commands; instead, they function as analysts who must evaluate uncertainty, calibrate trust, and determine when to rely on AI recommendations versus when to intervene \cite{b2} as illustrated in Fig. \ref{fig:pre_post_loop}. Consequently, human judgment and AI outputs must be viewed as complementary components of a shared decision-making process rather than as replacements for one another \cite{b3}.

Traditional UX methods were designed for deterministic systems where specific inputs lead to predictable outputs. This legacy approach assumes a level of stability and minimal uncertainty that is fundamentally at odds with AI systems, which produce false positives, false negatives, and continuously adapt over time. When interfaces fail to communicate this inherent uncertainty or fail to support complex user decision-making, the result is often over-reliance, under-reliance, or operational confusion \cite{b4} \cite{b5}. In high-stakes domains—such as the video anomaly detection system explored in this work—these mismatches can have severe implications for safety, accountability, and the ethical allocation of resources \cite{b6, b25, b26}.

To address these challenges, AI systems require a deeper integration of human-centered principles. Human-in-the-Loop (HITL) approaches should not be viewed merely as technical safeguards; rather, they are core UX design strategies that empower users to apply domain expertise and maintain appropriate oversight \cite{b3} \cite{b2}. Effective HITL design supports trust calibration \cite{b4}, improves decision quality, and mitigates improper reliance \cite{b7}.

Furthermore, as AI becomes embedded in real-world workflows, its impact extends beyond the individual to the organizational level. System-level metrics and oversight practices influence institutional trust and long-term adoption. This broader integration moves the field from a Human-in-the-Loop focus toward a Society-in-the-Loop paradigm, where UX design must account for organizational constraints and regulatory expectations. In this context, user experience serves as the critical bridge between technical system performance and societal integration.

In this paper, we make the following contributions:

\begin{itemize}
\item Conceptualize the shift in user experience (UX) design by distinguishing between the "operator-centric" pre-AI and the "analyst-centric" post-AI interaction paradigms.

\item Position Human-in-the-Loop (HITL) as a fundamental UX design strategy rather than a secondary technical safeguard.

\item Propose a set of post-AI UX evaluation dimensions, specifically operationalizing metrics for trust calibration, operational latency, and adaptation time.

\item Develop and implement an operational HITL architecture within a real-world AI-enabled surveillance system to validate these concepts.

\item Empirically analyze stakeholder perceptions across high-risk contexts to establish design principles for Society-in-the-Loop integration.

\end{itemize}

 \section{Related Works} 
As machine learning and autonomous decision-making integrate into interactive systems, UX has shifted from static interface design toward human–AI interaction. This transition, described as UX 3.0 \cite{b8}, reframes user experience as the need to interpret and calibrate trust in intelligent systems rather than merely operate software. Consequently, post-AI UX research centers on explainability, trust calibration, and human–AI collaboration.

Explainable AI (XAI) has become central to post-AI UX due to the opacity and autonomy of machine learning models. Prior work argues that explainability should be evaluated based on what makes explanations meaningful to users rather than merely transparent \cite{b9}. Research further shows that effective XAI must align with users’ goals and contextual needs \cite{b10}, and that explanations should support user reasoning instead of simply exposing model internals \cite{b11}. From a UX perspective, explainability therefore focuses on enabling sensemaking and informed interaction.

Trust, closely linked to explainability, concerns how users calibrate reliance on probabilistic AI systems. Prior research conceptualizes trust as a dynamic, experience-dependent construct shaped through interaction rather than a fixed trait \cite{b12}. It emphasizes the importance of appropriate trust, supported by transparency cues, feedback mechanisms, and uncertainty communication to prevent both over- and under-reliance. Trust-aware design therefore aims to align users’ perceived system capabilities with actual performance.

Adaptive and collaborative systems reflect a post-AI UX paradigm in which AI dynamically adjusts its behavior and participates in shared decision-making. Prior work describes human–AI collaboration as involving shared cognition, coordination, and evolving roles within workflows, positioning AI as an active participant rather than a passive tool \cite{b13}. From a UX perspective, such systems introduce challenges related to user agency, coordination, and understanding behavioral adaptation over time.

\begin{figure}[h!]
    \centering
    \includegraphics[width=0.8\columnwidth, trim=50 450 100 80, clip]{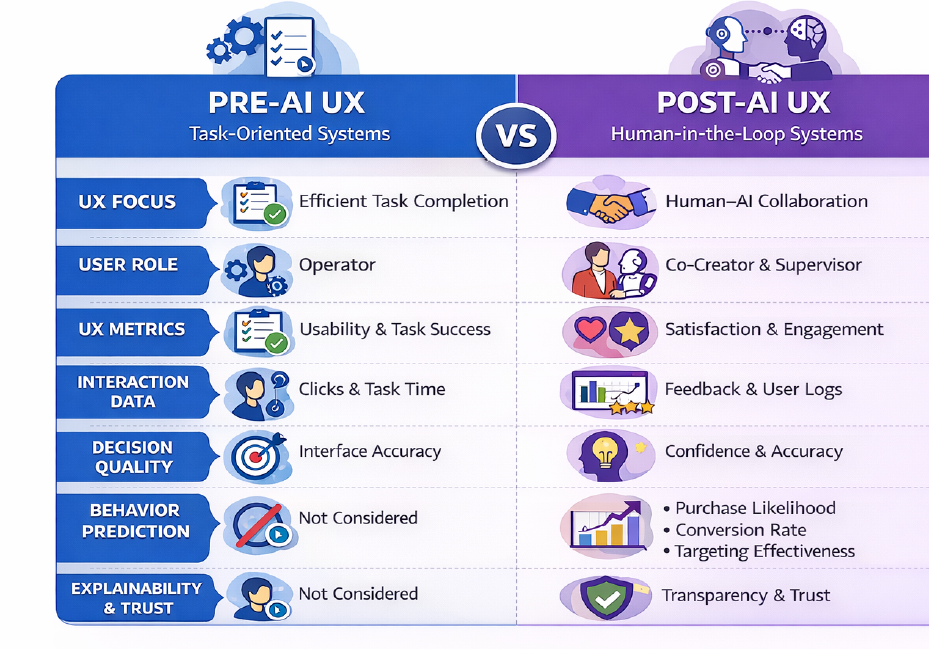}
    \caption{Comparing the Pre-AI UX focuses and the Post-AI UX Impications. }
    \label{fig:prepost}
\end{figure}

As AI systems become embedded in real-world workflows, UX research has moved beyond interface design to conceptualize human–AI interaction as an evolving ecosystem shaped by feedback and co-adaptation. This process, described as human–AI coevolution \cite{b14}, highlights how human and model behaviors continuously influence and reshape one another over time.
Fig.~\ref{fig:prepost} summarizes the transition from pre-AI to post-AI UX workflows. 

Within this perspective, Human-Centered AI (HCAI) promotes sustained human involvement across the AI lifecycle \cite{b15}, often implemented through Human-in-the-Loop (HITL) mechanisms. However, prior work largely treats HITL as a technical safeguard, with limited examination of its implications for user experience and long-term interaction..

\section{Pre-AI vs. Post AI User Experience}

% HITL commonly used for:
%Model validation
%Error correction (FPR /FNR)
%Safety oversight
Prior work has primarily treated Human-in-the-Loop (HITL) as a technical safeguard in AI systems. Human involvement is commonly framed as model validation and error correction \cite{b16}, while AI safety research emphasizes supervisory intervention to prevent harmful outcomes \cite{b17}. Together, these perspectives position HITL as a mechanism for improving correctness and safety rather than as a deliberately designed user experience

Despite its widespread adoption, Human-in-the-Loop (HITL) is rarely framed as a UX design strategy. It is typically treated as a mechanism for improving correctness and safety, positioning users as supervisors rather than experiential participants. Consequently, evaluations prioritize system performance while overlooking UX outcomes such as trust calibration and long-term interaction. This gap highlights the need to reconceptualize HITL as an experience-centered component of human–AI systems.

In response to this gap, we treat HITL not only as a model correction mechanism but as an experience-centered interaction paradigm embedded in real-world workflows. Although technically evaluated, the system is deliberately designed to support human agency, contextual judgment, and distributed oversight in line with human-centered and society-in-the-loop principles.

\section{Social Construction} 

\subsection{Analytical Framework and Inductive Theme Development\protect\footnotemark}
\footnotetext{The dataset used in this study is derived from \cite{b21}.}

Understanding the sociotechnical implications of AI systems requires more than technical evaluation. While prior sections discuss system design and performance, they do not fully capture how stakeholders interpret, experience, and react to these systems in real-world settings. Because this paper argues that the transition from pre-AI to post-AI UX reshapes organizational and social dynamics, it was necessary to examine how different stakeholder groups perceive analytics-based systems. This qualitative study was therefore conducted to provide empirical grounding for the Society-in-the-Loop framework proposed in this paper.

This study employed an inductive qualitative approach \footnote{This study was reviewed and approved by an Institutional Review Board; protocol details are withheld in accordance with the single-blind review policy.} to examine how diverse stakeholder groups articulate their needs, perceptions, concerns, and constraints regarding the use of analytics-based systems. Data were collected through focus groups and semi-structured interviews across multiple contexts, including universities, law enforcement agencies, detention facilities, and commercial retail environments.
Rather than analyzing each interview as a single unit, participant statements were segmented into smaller analytical units, each representing a distinct idea, concern, or expectation.
Through this process, 269 discrete insights were identified and analyzed individually. These insights were then organized using an inductive coding process that resulted in four overarching analytical categories: 
\begin{itemize}
    \item Security personnel
    \item Local business owners
    \item Law enforcement professionals\end{itemize}

\begin{figure}[h!]
    \centering
    \includegraphics[width=\columnwidth, trim=70 500 180 50, clip]{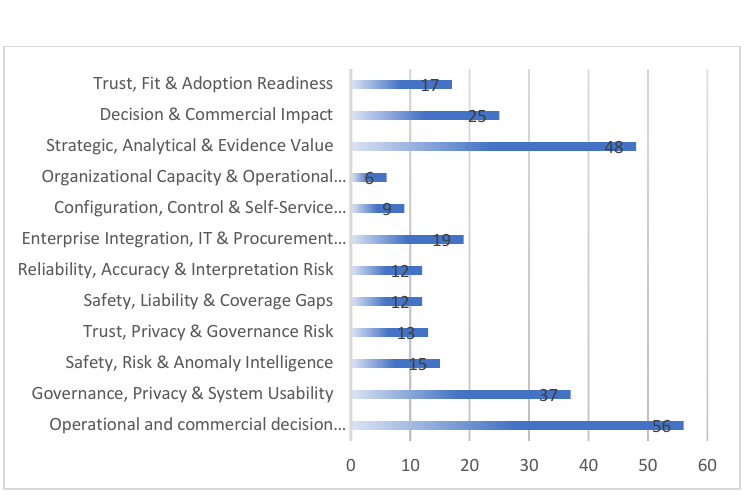}
    \caption{Distribution of Coded Insights Across Themes }
    \label{fig:dist}
\end{figure}
    
Fig. \ref{fig:dist} illustrates how these 269 insights are distributed across themes and analytical categories. The figure highlights the relative concentration of insights within each theme and provides an overview of how stakeholder input is spread across requirements, concerns, constraints, and perceptions. This distribution offers an empirical foundation for understanding which themes were most salient across stakeholder groups. The distribution of insights across themes shows that technical performance is closely connected to organizational concerns. Themes related to decision support, governance, reliability, and risk frequently overlap, suggesting that system accuracy and functionality cannot be separated from trust, liability, and institutional responsibility. This pattern supports our argument that AI system evaluation extends beyond technical correctness and directly influences social and organizational structures.

\subsection{Decision Support and Analytical Value of AI Systems}
Across stakeholder groups, analytics-based systems were primarily valued for their ability to support real-world decision-making. Participants emphasized that the most important function of AI systems was to assist operational, managerial, and business decisions rather than merely monitoring environments.

The most prominent theme related to decision intelligence in operational and business contexts. Stakeholders wanted systems that could help them make more accurate and informed decisions about staffing levels, operational scheduling, resource allocation, and the optimization of sales and service delivery.

A closely related theme was analytical and evidence-based value. Participants highlighted the importance of having numerical data, reports, and analytical outputs to support decisions that were previously based on intuition. These findings indicate that analytics systems were viewed as tools for strengthening decision quality rather than replacing human judgment.

\subsection{Risk Sensitivity, Errors, and False Positives in High-Stakes Contexts}
{Following initial insight extraction, an iterative thematic analysis was conducted. Insights were first reviewed independently and then grouped based on conceptual similarity. Through multiple rounds of refinement, all insights were organized into four overarching analytical categories:
\begin{itemize}
    \item {Requirements: Functional or operational needs stakeholders expect a system to fulfill}
    \item {Perceptions: Stakeholders’ beliefs, interpretations, and perceived value or relevance}
    \item {Concerns:  Risks, fears, or apprehensions related to adoption or use}
    \item{Constraints: Structural, organizational, technical, or regulatory limitations affecting implementation}\end{itemize}

One of the key patterns identified in the analysis was that as systems were perceived to generate greater value, stakeholder concerns also increased. When systems were expected to influence high-stakes decisions, participants expressed heightened sensitivity to potential errors, misinterpretation, lack of transparency, and false positives.

For example, participants noted that if normal behaviors were incorrectly classified as abnormal, security and risk-detection systems could become both highly useful and highly risky. False alarms were associated with potential legal, operational, and organizational consequences. These findings suggest that increasing system importance is directly associated with increasing concern over accuracy and error tolerance.

\begin{figure*}[h!]
    \centering
    \includegraphics[width=0.8\textwidth, trim=0 25 0 25]{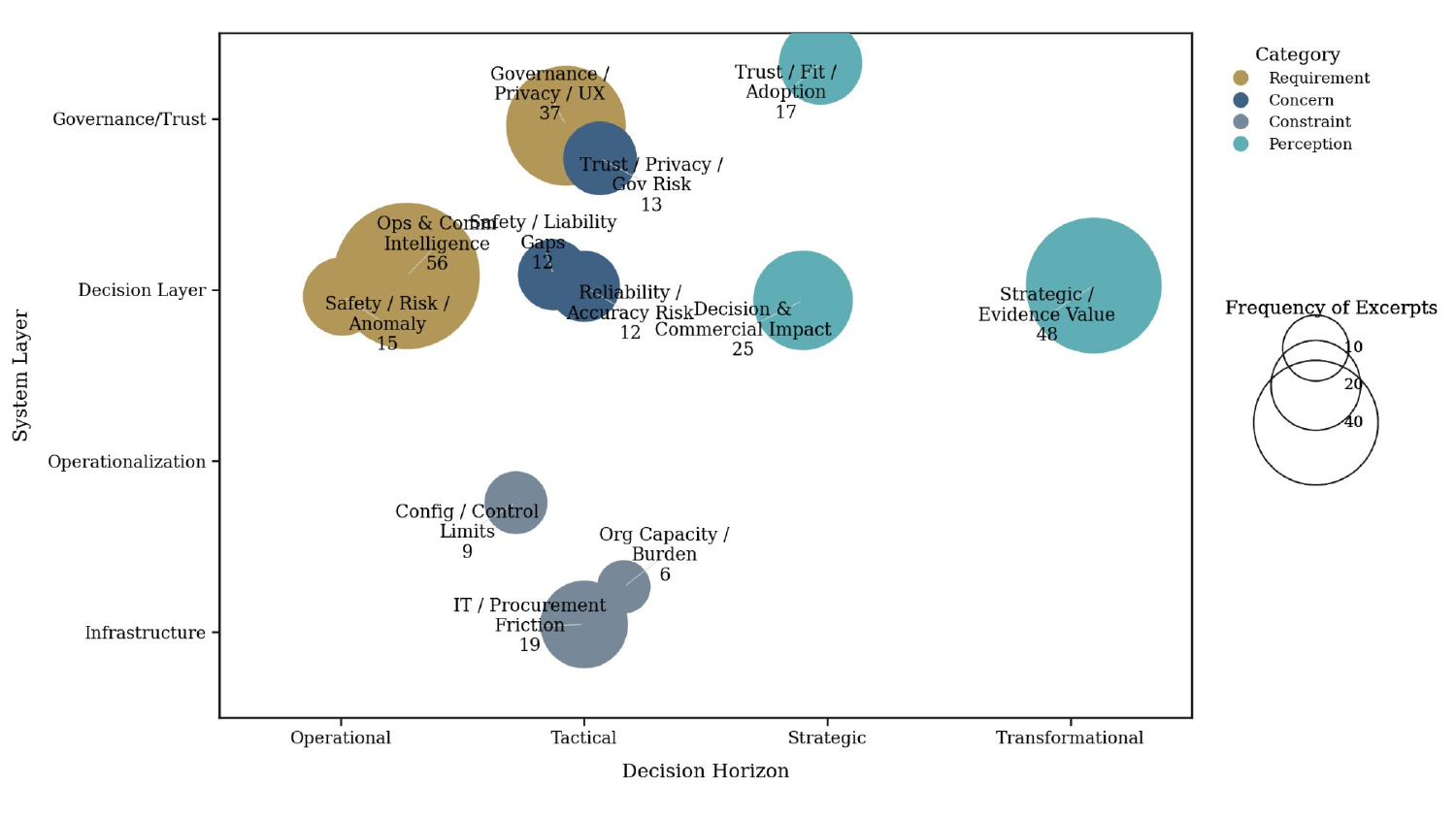}
    \caption{Theme Landscape by Decision Horizon and System Layer. Bubble size reflects theme frequency; color indicates category. Labels are automatically adjusted for clarity and connected to their bubbles.}
    \label{fig:Theme}
\end{figure*}

\subsection{Trust, Transparency, Privacy, and Governance}

Issues related to trust, transparency, and governance emerged as central conditions for system adoption. Participants discussed data privacy, regulatory compliance, and explainability not merely as formal requirements, but as essential foundations for trust.

Systems that lacked transparency or clear explanations for their outputs were viewed as difficult to trust, regardless of their technical sophistication. These concerns frequently intersected with decision-support themes, particularly in high-risk operational environments.

\subsection{Human-in-the-Loop and Organizational Control}
Participants consistently expressed a preference for systems that augment human judgment rather than replace it. Human-in-the-loop approaches were viewed positively, particularly when systems supported decisions while preserving human oversight.

Importantly, participants emphasized role-based control over system configuration. While they wanted transparency regarding how systems function and how alerts are generated, they did not support unrestricted access to system settings. Core configurations were expected to remain under the responsibility of designated individuals, while regular users focused on interpreting and reviewing outputs.

Figure \ref{fig:Theme} visualizes the overlap between themes across analytical categories. The bubble chart highlights how decision support, trust, risk sensitivity, and governance are not isolated concerns but frequently co-occur across stakeholder insights. This overlap underscores that the primary challenges of human-in-the-loop systems are not just technical. Instead, successful deploy depends on balancing analytical value, trust, and organizational control with a broader socio-technical context.

Taken together, these findings demonstrate that AI system performance metrics are not experienced as abstract technical numbers by stakeholders. Instead, they shape workload, responsibility distribution, risk perception, and long-term adoption. This empirical evidence reinforces the central claim of this paper: in post-AI systems, UX design operates within a sociotechnical ecosystem where technical design decisions influence organizational behavior and societal trust structures.

\section{Technology Experiment} 

\subsection{Human-In-The-Loop System Design} 
To address the contextual and subjective nature of anomaly interpretation in AI-enabled video surveillance, we add a Human-in-the-Loop (HITL) feedback layer to the baseline architecture. As shown in Fig.~\ref{HITL}, this design forms a closed loop between the AI pipeline and end users, enabling human oversight to refine detection performance while preserving user control. The system comprises four components—the AI Node, Server Node, Cloud Node, and End-User Interface—supporting secure data flow from sensing to feedback and continual learning. Access control restricts authorized users to site-specific data only.

At the AI Node, CCTV streams are analyzed to identify candidate anomalous events. Detected clips (5--10 seconds) are uploaded to an S3 bucket in the Cloud Node, avoiding local storage and supporting privacy-by-design. The upload triggers a Lambda function that notifies users via SNS and inserts a record into the DynamoDB Feedback table with a video URL, unique ID, and label field. AppSync exposes this metadata to end-user applications via GraphQL APIs for near real-time interaction. The interface (Fig.~\ref{HITL}, right) presents a binary validation prompt; responses are sent over HTTPS and stored as $+1$ or $-1$. As a single label per event was found sufficient, labeled clips are removed from other users’ queues. Validated labels are fed back to the AI Node for continual model updates, improving contextual sensitivity and reducing false alarms. Overall, the HITL design promotes human agency, accountability, and trust while strengthening generalization through context-aware feedback.
\begin{figure*}[htb!]
\centering
    \includegraphics[width=0.85\textwidth, trim=2cm 4cm 2cm 0cm]{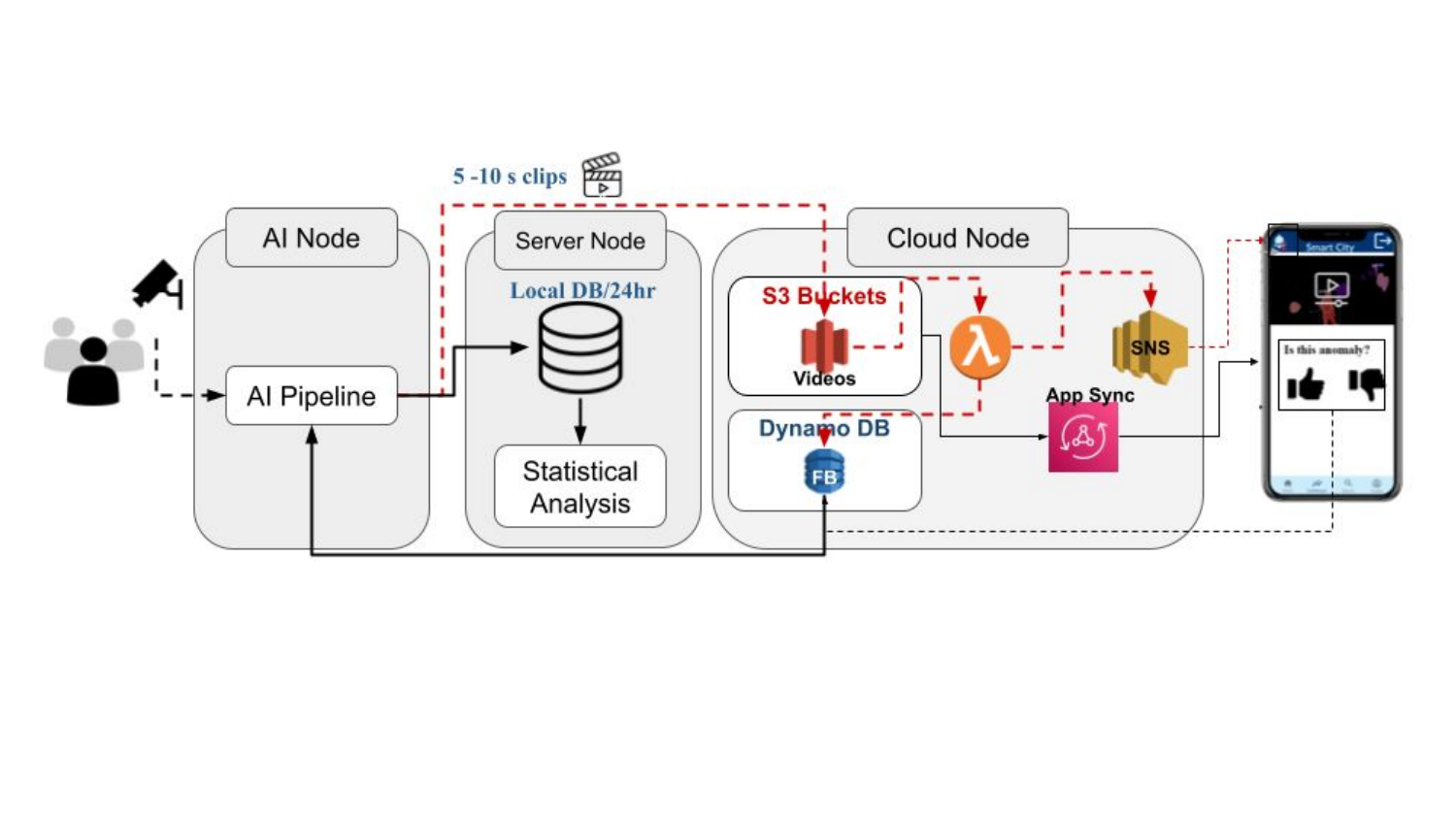}
\caption{End-to-end architecture of the proposed Human-in-the-Loop (HITL) system.}
\label{HITL}

\end{figure*}

\subsection{Event-Based Anomaly Detection}  
\begin{figure*}
\centering
    \includegraphics[width=0.8\textwidth, clip, trim=0 75 0 0]{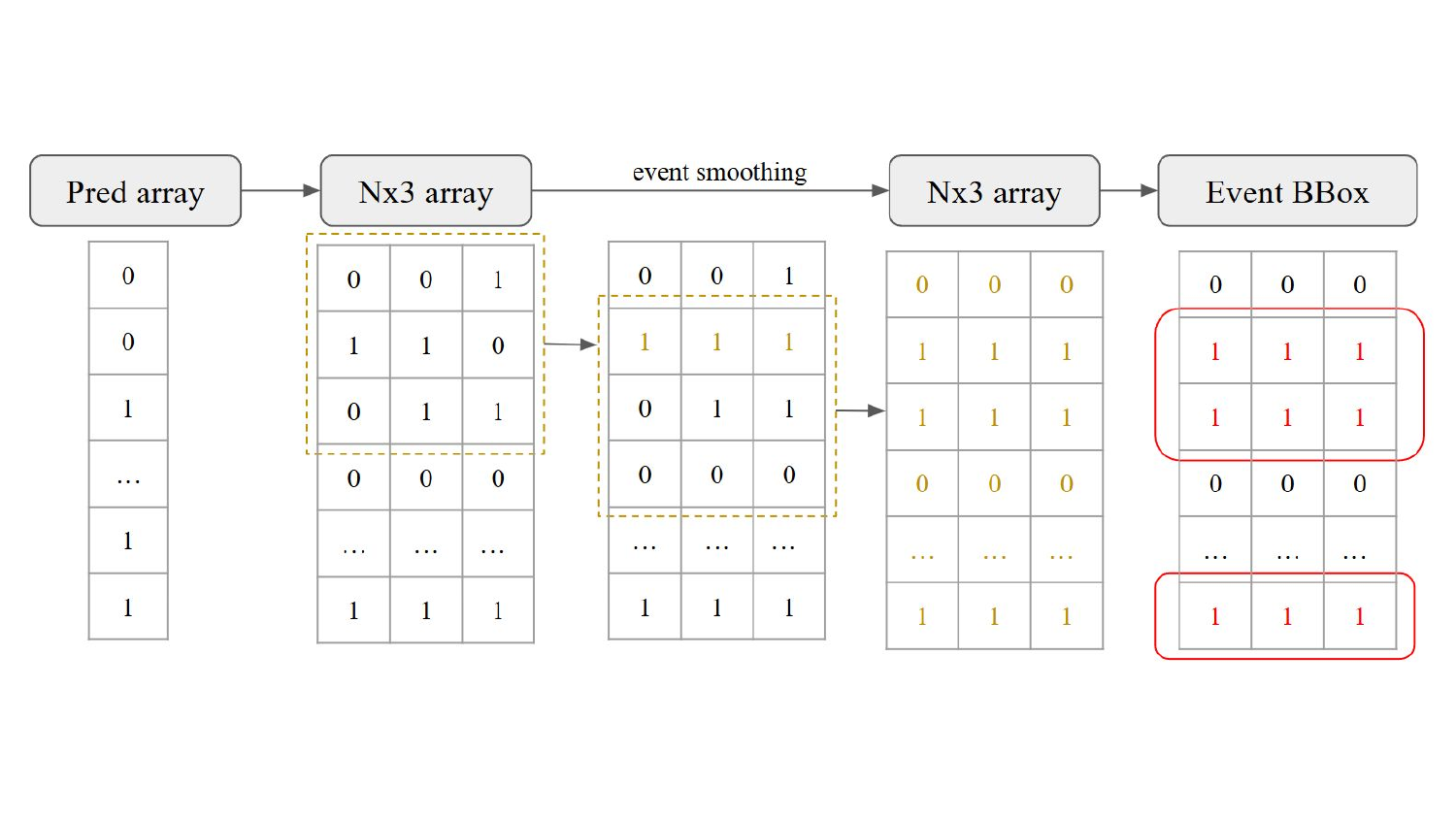}
\caption{From Frame-Based Anomaly to Event-Based Anomaly Detection Process.}
\label{Event}
\end{figure*}

The system uses STGNF for frame-level anomaly classification; however, frame-wise predictions often produce high false positives in practice \cite{b18}. Since anomalous behaviors are temporally coherent, we apply an event-based post-processing step to enforce temporal consistency and suppress spurious detections. As shown in Fig.~\ref{Event}, binary frame predictions are reshaped into an $N \times 3$ array and smoothed using a $3 \times 3$ sliding window. When more than $50\%$ of entries in a window are anomalous, the center row is set to all ones, with zero-padding applied at sequence boundaries. Contiguous anomalous segments are then merged into event-level intervals defined by start and end frames.

Event-level predictions are evaluated using IoU, with detections exceeding $0.5$ considered correct. As reported in Table~\ref{tab:event_detection}, the method achieves a precision of $0.731$ and recall of $0.750$ on the PoseLift test set \cite{b19}, demonstrating reduced fragmentation and improved detection stability over frame-level outputs.

\begin{table}[t]
\small
\centering

\caption{Event-based detection results evaluated on the PoseLift test set \cite{b19}.}
\label{tab:event_detection}
\scalebox{1.5}{%
\begin{tabular}{@{}l c@{}}
\hline
Metric & Count \\
\hline
GT events & 40 \\
Predicted events & 41 \\
TP\_detection & 30 \\
FP\_detection & 11 \\
FN\_detection & 10 \\
\hline
\end{tabular}
}
\end{table}

Traditional UX evaluation was designed for deterministic, task-oriented systems. In contrast, AI-enabled HITL systems create continuous feedback between users and the AI pipeline. Here, UX emerges not only from interface usability but from how users interpret outputs and observe system adaptation. Evaluating post-AI UX therefore requires attention to system behavior and feedback dynamics, not just interface design.

\begin{figure}[h!]
    \centering
    \includegraphics[width=0.7\columnwidth, trim=65 0 65 42, clip]{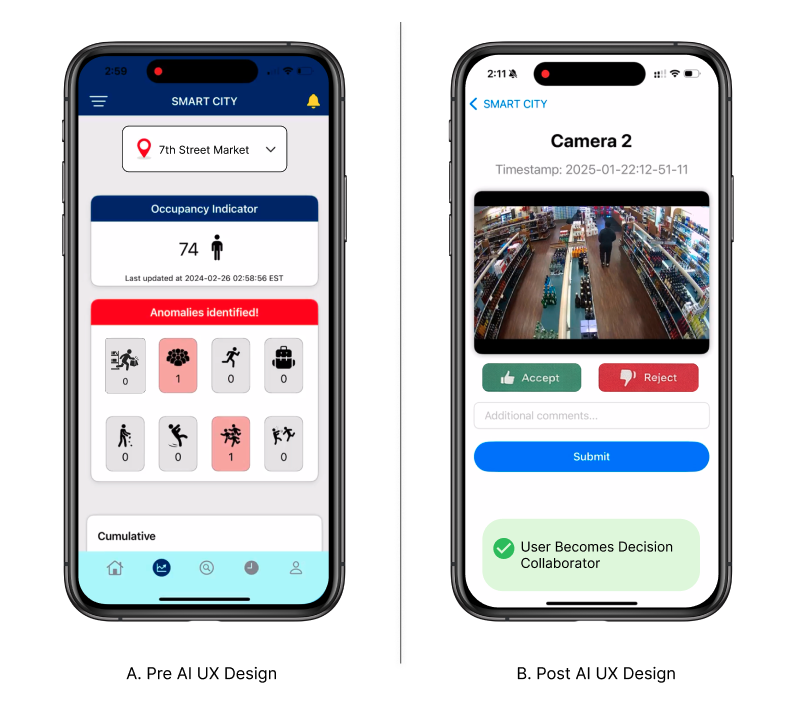}
    \caption{Comparison of UX Before and After Human-in-the-Loop Integrations. }
    \label{fig:UX}
\end{figure}

Within the proposed architecture, system-level performance, particularly false positives, has direct experiential consequences. Incorrect anomaly alerts require users to review clips and provide feedback, increasing cognitive effort and disrupting workflows. While retraining improves model performance, frequent false positives lead to alert fatigue. Reducing them therefore improves both detection accuracy and the sustainability of human oversight.

A second key dimension of post-AI UX is trust calibration. Users must decide whether to accept or override AI judgments, shaping trust through repeated interaction and perceived responsiveness. Design elements such as context-rich clips and simple validation support calibrated reliance. UX quality depends on informed intervention rather than blind trust.

Perceived controllability refers to users’ sense that their actions influence system behavior. In the HITL framework, users shape future detections through feedback used for retraining. UX quality depends on whether this influence feels meaningful. Simple mechanisms, such as binary feedback, reduce effort while reinforcing user agency.

Cognitive load management is a key UX concern in the HITL architecture. Users repeatedly interpret AI outputs and decide under uncertainty. Brief clips and simple feedback mechanisms reduce cognitive burden while preserving oversight, supporting sustained engagement in continuous monitoring contexts.

Post-AI UX in HITL systems is also shaped by human–AI alignment—the extent to which system behavior matches user expectations and contextual understanding. As feedback is incorporated into retraining, users may observe performance improvements over time. UX quality depends on whether these changes are perceived as consistent and interpretable, reinforcing users’ mental models. Unexpected shifts in behavior can undermine trust, even if technical performance improves. Figure \ref{fig:prepost}, summarizes this shift in UX focus before and after the rise of AI. 

Figure \ref{fig:UX} summarizes this transformation by contrasting the interaction model before and after Human-in-the-Loop integration. In pre-AI UX, the user primarily interacts with a deterministic system where outputs are fixed and system performance is evaluated separately from user experience. In post-AI UX, interaction becomes cyclical and adaptive. Users receive AI-generated outputs, provide feedback, and indirectly influence future system behavior. This shift illustrates how backend performance, feedback loops, and user oversight become integrated components of experience design. The figure therefore supports the paper’s central argument that UX in AI-enabled systems is no longer interface-centered, but ecosystem-oriented.

\section{Defining New UX Metrics}
As illustrated in Figure \ref{fig:prepost}, UX in AI-enabled Human-in-the-Loop systems extends beyond interface usability into broader sociotechnical structures. The social construction findings and bubble chart (Figure \ref{fig:Theme}) show that stakeholders experience AI through themes such as risk, reliability, governance, and organizational capacity indicating that UX now operates across operational and decision layers. Traditional UX metrics, developed for deterministic and interface-centered systems, do not capture these backend and organizational dimensions.
For this reason, we define a selective set of post-AI UX metrics grounded in the empirically derived themes. Related themes were grouped and merged into higher-level constructs to enable operationalization, including organizational capacity —comprising information flow and technical readiness, procurement friction, liability gaps, reliability, intelligence, control limits  and safety-risk. From this synthesis, four metrics emerge: Accuracy, Latency, Adaptation Time, and Trust. The following subsections define each metric.

\subsection{Accuracy}
Within the proposed post-AI UX framework, accuracy is defined not merely as a statistical performance metric but as a sociotechnical construct reflecting perceived intelligence and reliability. Intelligence refers to the system’s ability to detect meaningful anomalies, while reliability reflects behavioral consistency over time. In probabilistic AI systems, accuracy operationalizes these experiential dimensions.

Backend performance directly shapes user experience: false positives create interruptions and alert fatigue, while false negatives undermine situational awareness and trust. Accordingly, detection accuracy becomes a determinant of experiential quality.

Accuracy is measured using False Positive Rate (FPR) and False Negative Rate (FNR). In Human-in-the-Loop systems, reducing FPR lowers workflow disruption, and reducing FNR strengthens perceived intelligence. Thus, FPR and FNR function as both technical and UX metrics in post-AI environments.
\subsection{Latency}
Within the proposed post-AI UX framework, latency is conceptualized not merely as a technical performance indicator but as a socio-organizational construct linking AI-generated alerts to real-world action. It derives from themes of liability gaps and organizational information flow, reflecting how effectively alerts align with authority and decision rights.

When notification routing does not match responsibility structures, delays occur even if detection is timely. In Human-in-the-Loop systems, latency therefore becomes a UX metric, shaping perceived responsiveness and clarity of accountability.

Latency is decomposed into technical latency (incident to alarm) and organizational response latency (notification to action). This work focuses on organizational response latency, operationalized as the time between alert notification and recorded action using timestamped system logs. Lower latency indicates stronger alignment between information flow and responsibility, whereas prolonged latency suggests workflow or liability misalignment.

\subsection{Adaptation Time}
Within the proposed post-AI UX framework, adaptation time captures the temporal dimension of organizational adoption, how quickly an AI system becomes integrated into routine workflows. It derives from themes of organizational technical capacity and procurement friction, which together shape infrastructure readiness, deployment feasibility, and acquisition constraints.

In post-AI settings, adaptation extends beyond learning a new interface to aligning infrastructure, access rights, and operational procedures. The speed of this alignment influences perceived feasibility and organizational acceptance, making adaptation time a UX metric rather than merely an implementation variable.

Adaptation time is operationalized as the duration between deployment and stable workflow integration, assessed through structured organizational surveys capturing perceived integration speed and deployment burden. This provides a measurable indicator of how effectively technical readiness and procurement processes enable AI adoption.

\subsection{Trust}
In proposed post-AI UX framework, trust is conceptualized as an experiential construct emerging from the themes of control limits and safety/risk identified in the social construction analysis. Stakeholders expressed concern not only about system accuracy, but about whether the AI operates within acceptable governance, security, and accountability boundaries. Trust therefore reflects users’ willingness to rely on the system under conditions of uncertainty.

Control limits refer to perceived boundaries of oversight and accountability, while safety/risk concerns relate to potential negative consequences of system behavior. When these boundaries are clear and risks are perceived as manageable, trust increases; ambiguity weakens reliance. In this sense, backend governance structures directly shape user experience..

Trust is operationalized using validated trust-in-automation survey instruments administered as multi-item Likert-scale questionnaires. Established scales, such as the Trust in Automation Scale, demonstrate strong reliability and predictive validity across AI domains [22]. Building upon these validated frameworks ensures rigorous quantitative measurement while maintaining theoretical continuity with prior research.

\section{Discussion and SocioTechnical Implications}

The shift from pre-AI user experience to post-AI user experience is not only a change in interface design, but a change in how intelligent systems connect to social and organizational structures \cite{b1}. In pre-AI systems, which were mostly deterministic, performance metrics such as accuracy, error rate, and latency were mainly technical measures used to evaluate system quality. These metrics did not directly affect user experience or organizational behavior.

However, in post-AI systems that are based on probabilistic models, continuous learning, and Human-in-the-Loop mechanisms, the boundary between the technical layer and the experiential layer becomes blurred. In these systems, model performance, feedback mechanisms, output presentation, and human control all shape user experience at the same time and can also influence patterns of interaction and organizational behavior.

For example, metrics such as false positive rate, false negative rate, latency, or model adaptability are no longer just technical numbers. Each alert may require users to spend time checking it, making a decision, or providing feedback \cite{b7}. These metrics directly affect user workload, cognitive effort, and long-term engagement with the system. In this sense, performance metrics become experiential and social variables.

From a sociotechnical perspective, this means that UX design is no longer limited to ease of use or visual design. It also includes designing mechanisms for trust, responsibility distribution, transparency, organizational control, and human–AI alignment. In this framework, Human-in-the-Loop is not only a model correction layer, but a mechanism to regulate trust, maintain human oversight, and integrate the system into institutional structures.

As a result, the shift from pre-AI to post-AI UX represents a transition from the user as an operator to the user as a collaborator and supervisor within an intelligent ecosystem. This change affects not only interaction patterns, but also organizational adoption, long-term sustainability, responsibility distribution, and the integration of AI into individual and collective decision-making processes \cite{b24}.

\section*{Acknowledgment}
This research is supported by the National Science Foundation (NSF) under Award Number 2527312.

\end{document}